# Effective Non-Iterative Phase Retrieval of 2-D Bandlimited Signals with Applications to Antenna Characterization and Diagnostics

Giada M. Battaglia, Andrea F. Morabito, Roberta Palmeri, and Tommaso Isernia

*Abstract*—The Phase Retrieval problem is dealt with for the challenging case where just a *single* set of (phaseless) radiated field data is available. In particular, even still emulating the solution of crosswords puzzles, we provide decisive improvements over our recent approaches. In fact, by exploiting bandlimitedness and a suitable set of intersecting curves, we definitively lower the computational complexity (thus eliminating drawbacks) of our previous techniques. Numerical examples, concerning applications of actual interest, support the given theory and confirm the effectiveness of the developed procedure.

## I. INTRODUCTION

AS extensively discussed in the related papers [1],[2], as well as in an extensive literature (see for example [3]-[21]) the so-called 'Phase retrieval' (PR) problem is of interest in very many different branches of applied sciences, including antennas. In such a kind of problems, one wants to retrieve a complex function from measurements of its square amplitude distributions plus some additional a-priori information.

In antenna problems, the function to be retrieved could be (a component of) the radiated field, and the a-priori information is the support (and hence the location and dimensions) of the source generating such a field. Because of the Fourier based relationship amongst the source and the far-field for both cases of discrete (i.e., array) and continuous (aperture) sources, particular attention has been devoted to the case where measurements are taken in the Fraunhofer zone. Consequently, the signal of interest can be considered bandlimited, with a bandwidth related to the dimension of the source. Interestingly, by using the concept of 'reduced radiated field' introduced by Bucci and co-workers in [22],[23], near-fields can also be considered bandlimited provided suitable auxiliary variables (depending on the kind of source and measurement surfaces, as well as on their distance) are introduced.

In our recent contributions [1],[2], we have tackled the PR problem by introducing a new point of view emulating the solution of crosswords puzzles. In a nutshell, the idea was finding the multiplicity of solutions admitted by the 1-D PR problems along straight lines or concentric rings (which can be done by means of the Spectral Factorization (SF) technique [24]), and then prune the tree of all possible combinations along the different lines by means of congruence arguments. In particular, in [1], which focuses on the case of array antennas, the idea is solving for the field behavior along rows and columns (and eventually diagonals), and then using crossing points in order to discriminate amongst admissible or non-admissible field solutions along the different straight lines. As a relevant drawback, the computational complexity of the procedure grows very rapidly with the dimensions of the source.

The problem has been partially overcome in [2] (where continuous planar sources are dealt with) where 1-D PR problems are supposed to be solved along diameters and concentric rings. Then, initialization of the procedure using the smaller rings (where 1-D PR problems have just a few ambiguities), as well as some overlooked properties of the fields (and hybridization with [24]) allow the consideration of much larger sources, including the case where phaseless measurements are affected from noise. On the other side, one still needs to consider 1-D PR problems along diameters, so in case of larger and larger sources the corresponding 1-D PR problems have a huge number of possible solutions, with the inherent difficulties in discriminating amongst all of them.

In this contribution, we eliminate such a drawback by considering a third possibility dealing with (intersecting) curves all having a small length, and hence, by virtue of the bandlimitedness property, corresponding to fields having a small amount of variability. As a consequence, the corresponding 1-D PR problems will have a limited number of possible solutions, hence making much easier the crosswords processing and consequently the overall PR procedure.

Note that the chance we are pursuing, i.e., getting the actual 2-D complex field by a single set of phaseless data, is based on theoretical uniqueness results arising from the fact that 2-D polynomials (but for a zero-measure set) are not factorable. This is indeed deeply different from the corresponding 1-D problem. In fact, in such a case the field and the square amplitude distributions can be expressed as 1-D trigonometric polynomials, and the SF of the data can lead, through a 'zero flipping' procedure, to a large number of different complex fields all corresponding to the same power pattern (see [1],[2], [24] for more details).

Notably, in both the 1-D and 2-D cases, attention still must

Corresponding author: Tommaso Isernia (tommaso.isernia@unirc.it)

be paid to the so-called 'trivial ambiguities' affecting any PR problem [21], i.e.:

i) a constant phase on the spectrum;
ii) a linear phase on the spectrum;
iii) a conjugation of the spectrum;
iv) any combination of the above.

In fact, all modifications *i)-iv)* of the spectrum give raise to the same power pattern.

As far as the strategies to remove the above ambiguities are concerned, *(i)* corresponds to the same constant phase on the aperture source and can be fixed by choosing a phase reference, while *(ii)* implies a translation of the source and hence can be fixed by knowing the minimal circle enclosing the source itself[1]. Finally, the ambiguity *(iii)* corresponds to a reversal plus a conjugation of the source and can be dealt with by exploiting a-priori information about the source or the field, e.g., the support of the source or the phase of the spectrum in a few points. It is worth noting that, whenever such a-priori information is not available, the set of possible solutions arising from *(iii)* is reduced to just two spectra which are complex conjugate each of the other.

The paper is organized as follows. In Sections II and III we respectively present the optimal field representation and the basic idea enabling the proposed PR approach. Then, in Sections IV and V the solution procedure is respectively introduced and tested through different numerical experiments. Conclusions follow.

## II. FIELD REPRESENTATION ALONG (NON-CONCENTRIC) RINGS

In [2] we have given an analysis of the properties and possible representations of the fields along (concentric) rings in the spectral domain. In summary, we argued, on the basis of suitable expansions and bandlimitedness, that along any circle of radius $\bar{k}$ centered in the origin of the spectral domain one needs $(2\pi\bar{k})/(\lambda/2)$ samples, which are supposed to be uniformly spaced in the angular variable (for more details, see [2]). Then, by turning the Dirichlet kernel-based sampling representation [25] into a Fourier series, an accurate representation is given by:

$$F(\bar{k}, \phi) = \sum_{\ell=-H}^{H} C_\ell(\bar{k}) e^{j\ell\phi} \quad (1)$$

wherein $H = \bar{k}a$ and $a$ is the radius of the circular support. Notably, a similar expansion holds true for the square amplitude distribution, the trivial difference being a doubling of the summation indices. Therefore, the square amplitude distribution data, say $M^2(\bar{k}, \phi)$, can be conveniently represented as [24]:

$$M^2(\bar{k}, \phi) = \sum_{\ell=-2H}^{2H} D_\ell(\bar{k}) e^{j\ell\phi} \quad (2)$$

where $D_\ell$ is a Hermitian sequence.

Then, a key point of the procedure was expressing both (1) and (2) as the restriction to the unitary circle of a polynomial in the *z-variable* as:

$$F(\bar{k}, \phi) = \sum_{\ell=-H}^{H} C_\ell(\bar{k}) z^\ell \quad (3)$$

$$M^2(\bar{k}, \phi) = \sum_{\ell=-2H}^{2H} D_\ell(\bar{k}) z^\ell \quad (4)$$

so that, when $z = e^{j\phi}$, one turns back to (1) and (2).

In fact, such a circumstance and the fundamental theorem of algebra [26] allow the factorization of (4), and hence, by using the properties of the zeroes, the extraction of all the different possible expressions (1), (3) for the field.

Notably, the order of the trigonometric polynomials (3), (4) is smaller and smaller for decreasing values of $\bar{k}$, which allowed in [2] a solution procedure much better than in [1].

Now, the very simple circumstance we rely upon herein is that bandlimitedness is a global property of the spectrum. Hence, representations (1)-(4) still hold true along rings which are not centered in the origin provided coefficients depend on the circle at hand, and $\bar{k}$ is the radius of such a ring[2]. Such a circumstance provides definite advantages, eliminating the computational burden drawbacks of [2] in all steps of the crosswords-like processing way of thinking we had introduced.

## III. THE BASIC IDEA, AND A POSSIBLE SYSTEM OF CIRCLES

Given the above property, the simple idea we rely herein is to exploit a system of concatenating rings. These latter should be such:

a) to cover the visible part of the spectrum;
b) have a number of intersection points allowing for effective discrimination;
c) have small radius (to deal with low order polynomials).

Obviously, very many different choices are possible. In the following, as we deem it very convenient, we take inspiration from the so-called 'four-colors' covering of the earth in communication from satellites [27]. In such a scheme (see Fig. 1) the region of interest is first partitioned into a series of hexagonal cells giving rise to a honeycomb structure. Then, a pencil beam (with a circular footprint) can be associated to each cell in that application.

---

[1] In case of periodic signals, which is the case for signals defined on a circle, the only admissible linear phases correspond to an integer shift of the Fourier harmonics of the signal.

[2] A simple way to prove that (3) and (4) hold true whatever the circle at hand is the fact that the fields (spectra) on a generic ring could be thought as positioned on (translated from or to) a circle centered in the origin by using suitable linear phases on the source aperture. Hence representations from [2] exactly apply.

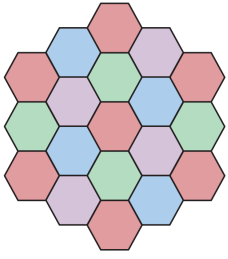

Fig. 1: Honeycomb structure used is satellite communications.

By paralleling such a situation, in our case, for each hexagon, we can consider the corresponding circle passing through the vertices. As a result, we end up with the system of circles pictorially depicted in Fig. 2. It can be note that one has a number of points where three different rings intersect (corresponding to the so-called 'triple points' in the literature [28]). As we discuss in the next section, such a circumstance allows for a convenient triggering and development of the proposed procedure.

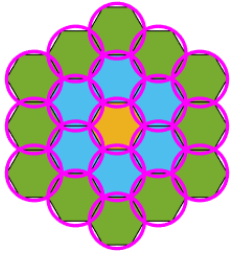

Fig. 2: Pictorial representation of the proposed system of circles for the coverage of the visible part of the spectral plane.

## IV. Triggering And Development Of The New Procedure

### A. Triggering of the procedure

By virtue of (1), (2) as applied to a generic circle of radius $\bar{k}$, one can trigger the overall procedure by considering two or even more (very) low order PR problems. For example, one can consider the three rings of Fig. 3, and choose the (identical) radius in such a way that $H$ is the minimum number such to satisfactorily fit the available data, i.e., to satisfy within some given tolerance the following expression:

$$\sum_{\ell=-2H}^{2H} D_\ell(\bar{k}) z^\ell = M^2(\bar{k}, \phi) \quad (5)$$

As it can be seen, the three rings have a common point, $P_0$, which will be used as a reference for phase normalization (e.g., for the choice of the phase reference), and three intersection points amongst two of the circles ($P_{01}, P_{02}, P_{12}$).

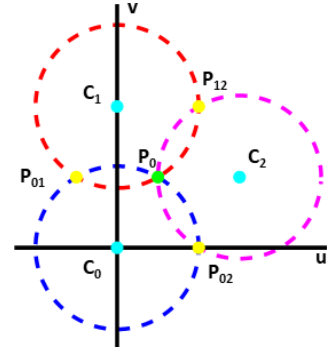

Fig. 3: Pictorial representation of the proposed procedure: triggering. The (signals value in) point $P_0$ (green marker) is used to set a reference phase for all the three signals along the circles, while the (signals value in) points $P_{01}, P_{02}, P_{12}$ (yellow markers) are used as discrimination points to discard solutions. In this configuration: $C_1 = \sqrt{3}\bar{k}\left(\cos\left(\frac{\pi}{2}\right), \sin\left(\frac{\pi}{2}\right)\right)$, $C_2 = \sqrt{3}\bar{k}\left(\cos\left(\frac{13}{6}\pi\right), \sin\left(\frac{13}{6}\pi\right)\right)$, $C_0 = (0, 0)$.

These latter will be used for discriminating amongst acceptable or non-acceptable field solutions along the rings[3].

If $H = 1$, and no zero is present within the ring(s) at hand, one just has two possible field solutions along each ring [2], so that a total of 8 possible configurations have to be checked. Such a number grows to 64 for $H = 2$, and to 512 for $H = 3$, which is anyway much less than the minimal number of combinations to be checked with our previous choices [1],[2]. Moreover, in most cases one will not really need to explore all of them, as each intersection point will provide a pruning of the set of possibilities. For example, the first intersection in point $P_{12}$ (requiring to explore 4, 16 or 64 possibilities at the intersection points when $H=1$, 2, or 3, respectively) will already reduce the set of overall possibilities, and the same will apply at the two other intersection points.

Finally, as detailed in [2], if the spectrum inside the circle has no zeroes, the $2H$ zeroes are located half inside and half outside the unitary circle of the complex plane, and one can reduce the number of combinations to be checked along each ring from $2^{2H}$ to the binomial coefficient $\binom{2H}{H}$, allowing significant improvements in performance.

Going into details, to trigger the procedure[4] the steps summarized in the flowcharts in Figs. 4 and 5 can be pursued. By assuming to start with the ring $C_0$, let us denote by $\{S_0^j\}_{j=1,\dots,N_0}$ the $N_0$ different solutions along such a ring. Then, for each of the $N_0$ solutions, we proceed according to the description in Fig. 4. At the end of this step, we have a number $N_1$ of admissible couples of partial solutions, say $\{S_0^i, S_1^i\}_{i=1,\dots,N_1}$ (see Fig. 4). Note $N_1$ may be greater, equal or (hopefully) smaller than $N_0$. In fact, when more than one solution on $C_1$ is compatible with a solution on $C_0$, the number of possibilities increases, whereas whenever a solution on $C_0$ has no match with solutions on $C_1$, it is dropped, thus

---

[3] Of course, many other choices are possible for the initial setting.
[4] The 'triggering' starts after all the possible trial solutions along the three circles $C_0$, $C_1$ and $C_2$ have been computed, and the constant phase ambiguity by enforcing that all of them have the same phase in $P_0$ have been solved.

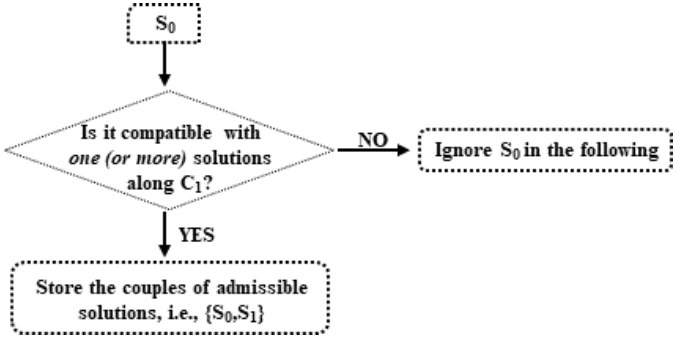

Fig. 4: Flowchart of the elementary bricks of step 1 of the proposed procedure. The routine must be repeated for each admissible solution $S_0$ along $C_0$.

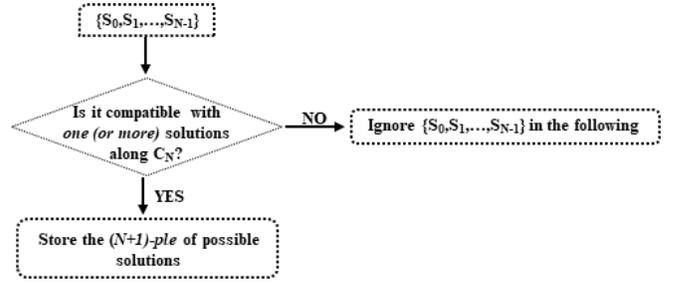

Fig. 6: Flowchart of the elementary bricks of step N of the proposed procedure. The routine must be repeated for each admissible tuple of solutions $S_0, S_1, \ldots, S_{N-1}$.

negatively contributing to the overall number of possible solutions.

Next, for each of these admissible couples, we enter in the flowchart of Fig. 5. At the end, we have a number $N_2$ of possible triplets of partial solutions, say $\{S_0^i, S_1^i, S_2^i\}_{i=1,\ldots,N_2}$.

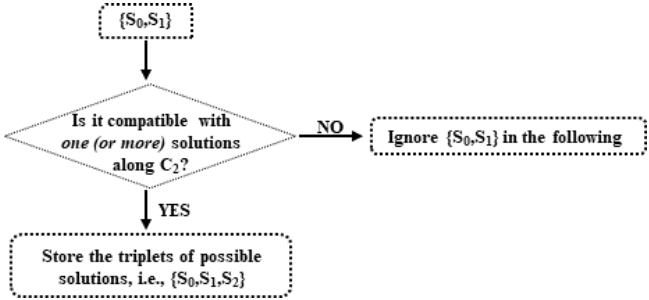

Fig. 5: Flowchart of the elementary bricks of step 2 of the proposed procedure. The routine must be repeated for each admissible couple of solutions $S_0, S_1$ along $C_0, C_1$.

Once again, depending on how many solutions are added (which happens when more solutions on $C_2$ are compatible with a single couple of solutions along $C_0$, $C_1$) or dropped (which happens when a couple $\{S_0, S_1\}$ has no compatibility with solutions on $C_2$), $N_2$ may be larger, equal or (hopefully) smaller than $N_1$.

*B. Development of the procedure*

The prosecution of the procedure is conceptually simple. In fact, one can consider additional concatenating rings passing through points where field reconstruction has already been achieved and implementing for each of them the procedure summarized in the flowchart of Fig. 6. For example, by referring to Fig. 7, one can proceed along the blue hexagons, and then progressively consider the hexagons with the next color, and so on. Note that for hexagons belonging to the same 'color' group, at each additional hexagon/circle (but for the last one) one has three intersections so that one of them can be used to fix the proper phase constant and the other two will allow for possible discrimination and hence pruning of the set of tentative solutions (see Fig. 7).

Note also that the last hexagon (circle) of the 'color' group allows indeed for four intersection points with the already

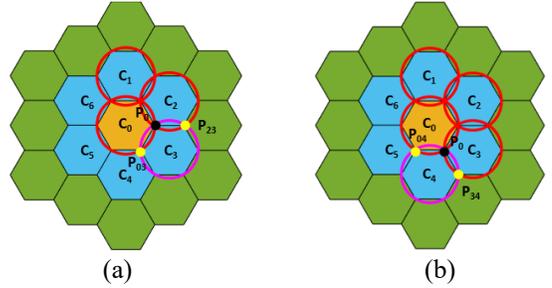

Fig. 7: Pictorial representation of the rationale of the proposed strategy. (a) and (b) represent two consecutive steps. Along red lines the complex field has already been retrieved, while in the magenta line has yet to be recovered. The field value at $P_0$ (black marker) is used to set a reference phase for all the signals along the circles, while the field values at $P_{03}$, $P_{23}$ in (a), or $P_{04}$, $P_{34}$ in (b) (yellow markers) are used as discrimination points to discard solutions.

considered circles, which allows for better discrimination/pruning capabilities.

Finally, after the actual (plus its conjugate) solution will be available for any hexagon (circle) covering the visible part of the spectrum, some kind of interpolation (or a fitting of the optimal representation given by expression (22) of [2] to the spectrum) will conclude the procedure.

For a better understanding of the procedure, we report in Fig. 8 a possible development of the general solution procedure assuming, by the sake of simplicity, that just four rings, i.e., $C_0$, $C_1$, $C_2$, $C_3$, allow the correct identification of the actual solution (but for the complex conjugate one). Under this assumption, the exemplifying tree of Fig. 8 corresponds to the following evolution of the process:

1) (Level 0) supposing $C_0$ is the starting ring, the SF technique is applied to find all $N_0$ possible solutions along $C_0$.
2) (Level 1) for each of the $N_0$ solutions, check for congruence between the trial solution along $C_0$ and the possible solutions along the second ring, (i.e., $C_1$) according to the procedure of Fig. 4. In particular, congruence is checked on the basis of the unwrapped phase misfit[5]. In the illustrative tree of Fig. 8, only 3 of the level zero solutions have some congruence with

---

[5] Note one does not require exact equality to accommodate discrepancies due to measurement errors [1],[2].

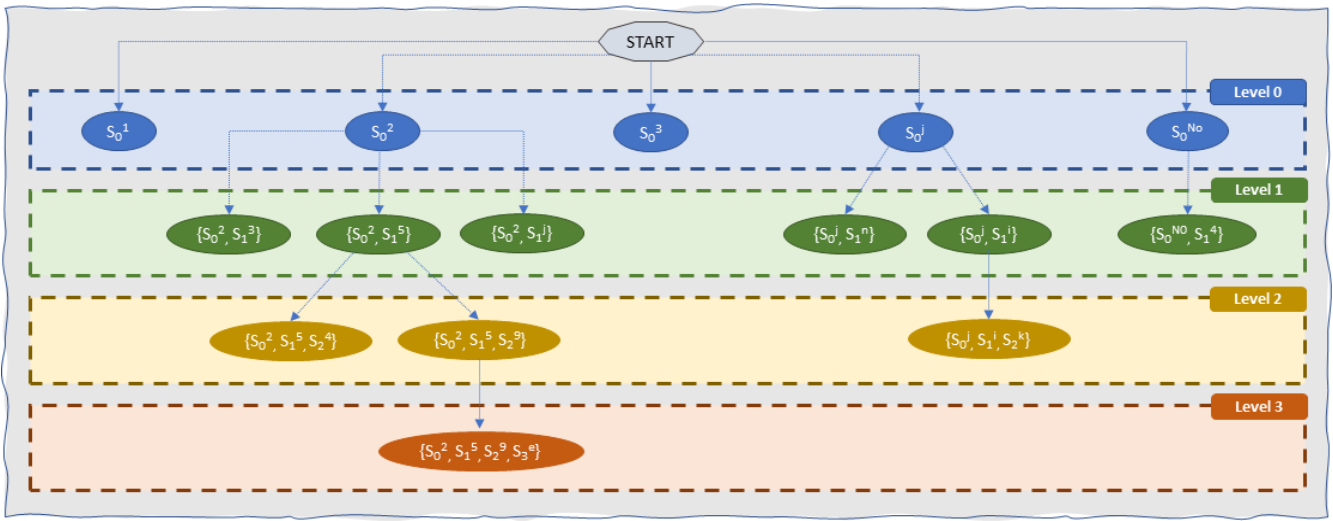

Fig. 8: A possible tree of tentative solutions.

   admissible fields on ring $C_1$, and they give rise to $N_1=6$ couples of solutions.
3) (Level 2) following Fig. 5 and the same misfit test as above, check the compatibility of the (level 1) $N_1$ tentative solutions with the possible fields on $C_2$. In the given virtual example, one comes out with $N_2=3$ possible triplets of solutions.
4) (Level 3) by iterating the procedure, a single solution is found at level 3 (and beyond) of the exemplifying tree at hand. Obviously, depending on data, the 'tentative solutions' tree may be much more cumbersome.

*C. Implementation details*

Two key aspects of the procedure are the data extraction along the non-concentric rings, and the choice of the radius $\bar{k}$.

The first issue arises from the fact that, to solve the PR problem along different non-concentric rings, one needs a square-amplitude representation of the data as prescribed by equation (4). This is a necessary condition to apply the SF theory [24] and find the different candidate solutions along each ring. To this end, one can compute the actual square amplitude samples in $4H+1$ equispaced points along the ring, and then perform a Discrete Fourier Transform (DFT), which will generate the required $D_p(\bar{k})$ values. As far as the computation of the samples is concerned, different methods can be used.

As a first and more obvious choice, if $(u_d, v_d)$ is the point of the spectral plane where the value of $M^2$ is required, one can rely on the explicit computation of the cardinal series representation. As no Fast Fourier Transform (FFT) is used, such a choice can be computationally intensive in case of larger and larger sources.

A faster procedure takes advantage from the effectiveness of FFT codes. In fact, starting from the initial samples and using FFT, zero padding and inverse FFT, one can get an interpolation of the $M^2$ distribution on a much denser grid. In such a way, one is able to make the power pattern available at points as close as possible to the desired ring. Then, by using the points of the grid closest to the ring (and neglecting the non-null distance from the ring itself) one can proceed to a best fitting procedure amongst representation (4) (with $z = e^{j\phi}$) and the values of $M^2$ on closest points. Notably, a sufficiently dense grid and the relatively low value of $H$ (which allows for filtering high frequency errors) allow to neglect the inherent approximation error.

A third intermediate possibility, starting again from the Nyquist grid samples, amounts computing the field (spectrum) on a denser grid by using FFT based interpolation. Then, one of the self-truncating sampling series of [29] could be used, which for any desired $(u_d, v_d)$ point allows getting the required value using a summation over a limited number of nearby sampling points of the dense grid.

In the numerical analysis which follows, the second strategy is exploited. Note that, in all three methods truncation of the measurement domain in the spectral plane (as one just can measure the visible part of the spectrum) can imply an interpolation error. Such an error is anyway very small in the central part of the spectrum, and anyway small for directive (but not super-directive) sources. Moreover, it is absent in case of uniformly spaced arrays (and hence periodic spectra) provided the spacing is sufficiently large (see [30] for more details).

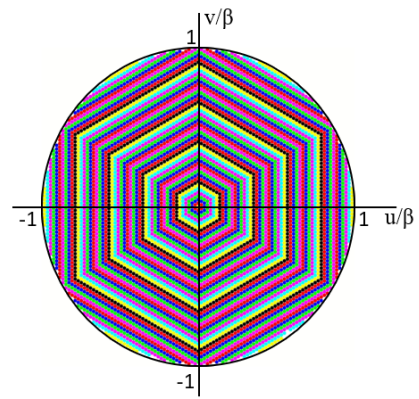

Fig. 9: Pictorial representation of honeycomb structure for the first numerical example.

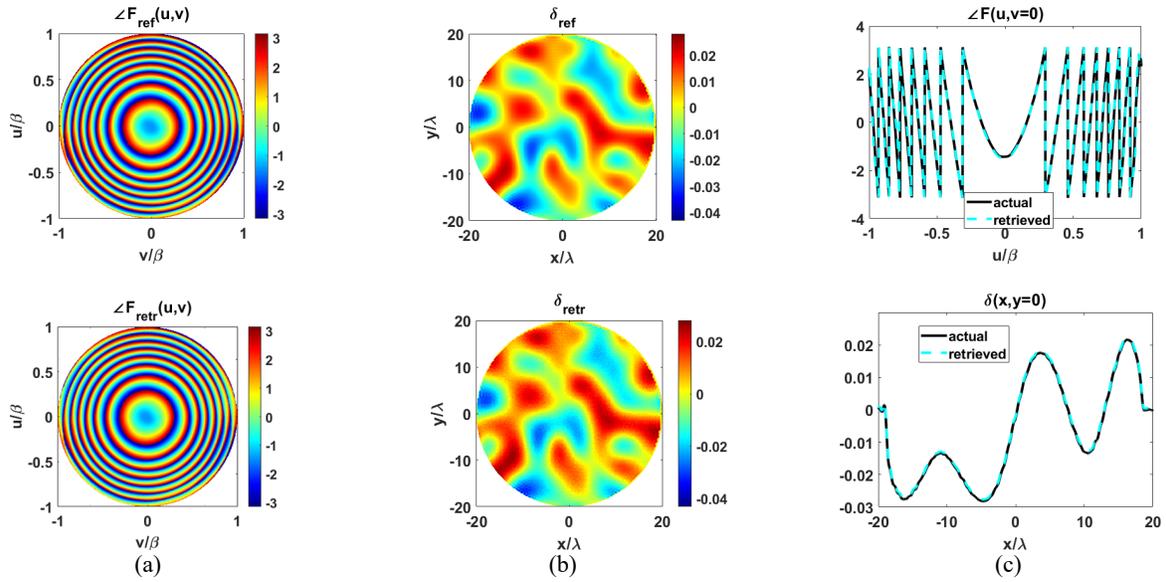

Fig. 10. Example #1: phase distortion on aperture field (with $u = k'\cos\phi$, $v = k'\sin\phi$). From left to right: (a) reference (top) and retrieved (bottom) phase of the radiated far-field; (b) reference (top) and retrieved (bottom) reflector surface deformation; (c) 1-D cuts of the reference (continuous black curve) and retrieved (dashed-cyan curve) far-field phase in $v = 0$ (top) and surface deformation along $y = 0$ (bottom).

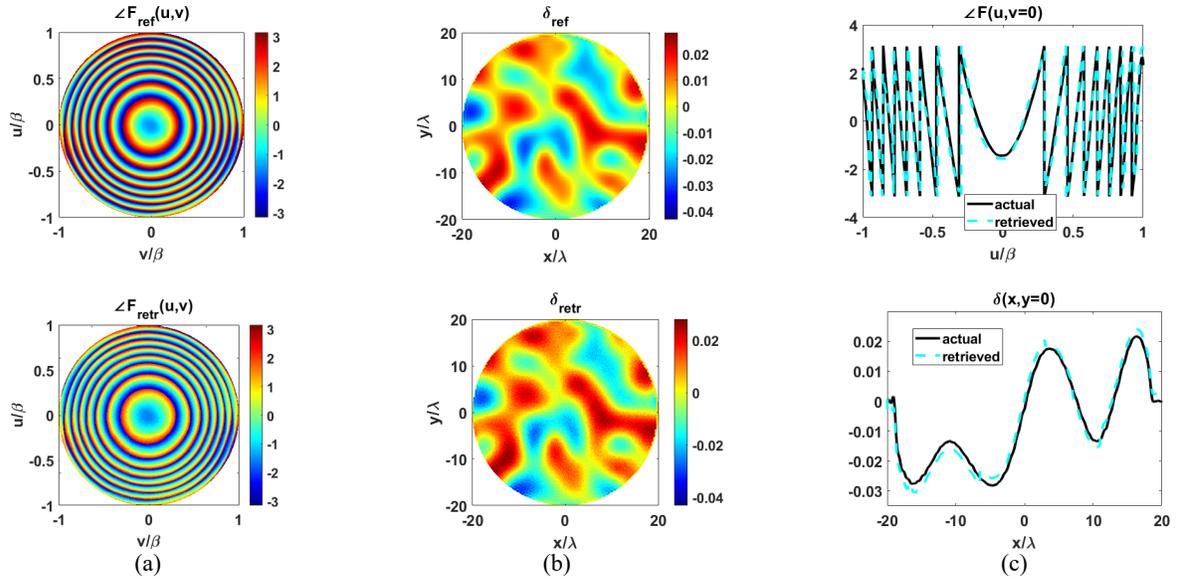

Fig. 11. Example #2 in presence of noisy data (SNR=25dB): phase distortion on aperture field. From left to right: (a) reference (top) and retrieved (bottom) phase of the radiated far-field; (b) reference (top) and retrieved (bottom) reflector surface deformation; (c) 1-D cuts of the reference (black curve) and retrieved (dashed-cyan curve) far-field phase in $v = 0$ (top) and surface deformation along $y = 0$ (bottom).

Notably, by virtue of the chosen approach, truncation of the invisible part of the spectrum will also be negligible by using a denser measurement grid (as compared to the Nyquist one) in the visible range, and self-truncating interpolation schemes.

A second general relevant issue concerns the choice of the radius $\bar{k}$. In fact, since a very small radius is suggested to deal with fewer combinations to be checked, fields at the different intersection points might not be independent one from each other, thus limiting discrimination capabilities. On the other hand, a larger radius ensures that fields at the intersection points will be independent, but at the expense of an increase of $H$ and, accordingly, of the computational cost. As a trade-off, we found it convenient to choose a ring radius equal to half of the Nyquist distance [2] (for the unknown spectrum).

### D. Possible Drawbacks

The proposed approach is affected by some limitations as well as from a potentially interesting characteristic.
If one has a null at an intersection point, phase makes no sense, so that one cannot use such a point for discrimination purposes. Anyway, apart from very peculiar cases, the other intersection points will allow for some discriminations. Also, one may consider a different ordering of the rings or even a different $\overline{k}$.

If the fields are factorable, the basic procedure furnishes in principle all the different solutions of the problem at hand, which could be a huge number. In fact, factorable fields represent the (zero-measure) set of 2-D cases which is affected by the same solution ambiguities related to zero flipping or the like one experiences in the 1-D case. Ambiguities due to the SF of the 2-D overall power pattern may come into play, leading to a high-number of complex 2-D fields all matching the measured amplitude data. Unfortunately, there is no remedy to such a problem which is very unlikely in the general case.

## V. NUMERICAL EXAMPLES

Several numerical examples have been performed to illustrate the effectiveness of the proposed approach. In the following, we focus on the diagnosis of surface deformations of a reflector antenna in Subsection V.A, while in Subsection V.B we deal with the retrieval of the excitations of a planar array. Note that, in both cases, we do not exploit any phase measurements of the field.

To perform a quantitative assessment, in each test case we evaluated the normalized square error metric for the radiated field (NSE$_{rf}$), defined as:

$$NSE_{rf} = \frac{\|F^{nominal}(u,v) - F^{recovered}(u,v)\|^2}{\|F^{nominal}(u,v)\|^2} \quad (6)$$

Moreover, based on arguments in Sect. IV.C above, $H = 4$ is adopted in all the following numerical examples.

### A. Diagnosis of surface deformations

In order to compare performances of the present approach with respect to [2], we consider herein the same kind of deformation.

In particular, the reference scenario is a continuous aperture field with a circular support of radius $a = 20\lambda$, $\lambda$ being the operating wavelength, that reads (see eq. (26)-(27) in [2]):

$$f(\rho', \phi') = |f|e^{j(\varphi_f + \Delta)} \quad (7)$$

$$|f| = \frac{4FL}{4FL^2 + \rho'^2} \quad (8)$$

$$\varphi_f = \beta \left[ 2FL + \tilde{C}\left(\frac{4FL^2 - \rho'^2}{4FL}\right)\right] \quad (9)$$

$$\Delta = \frac{8FL^2\beta}{4FL^2 + \rho'^2}\delta \quad (10)$$

wherein $\rho'$ and $\phi'$ are the radial and azimuth coordinates spanning the aperture, $FL$ represents the focal length, $\tilde{C}$ is a constant, $\delta$ corresponds to a surface deformation on the reflector, and $\beta$ indicates the wavenumber. Note that in (7) $\Delta$ is a space-dependent phase distortion [3] due to a surface deformation, that adds to the nominal phase $\varphi_f$, whereas the source amplitude $|f|$ keeps unaltered.

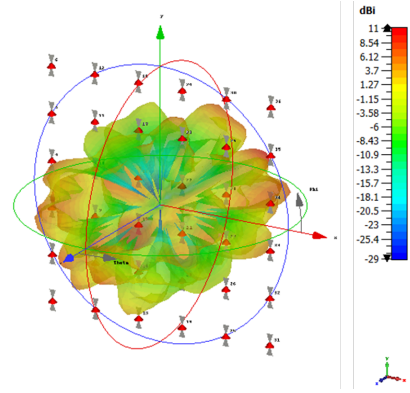

Fig. 12. Planar array of bow-tie antennas considered for the numerical assessment, with superimposed its directivity pattern (CST™ full-wave simulation).

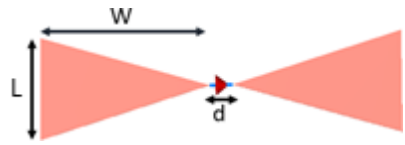

Fig. 13. Single radiated element for the planar array considered in the third numerical example and shown in Fig. 12.

In all the examples below, $FL=4\lambda$, $\tilde{C} = 0.5$, $\delta$ is randomly smooth in the range $\left[-\frac{\lambda}{30}, \frac{\lambda}{30}\right]$ as in [3], and a Gaussian taper has been superimposed to (8) to get an overall 20 dB ratio amongst values attained by the field at the origin and at the border of the disk source.

As far as the 2-D to 1-D PR problem decomposition is concerned, 6924 concatenating rings of radius equal to half of the Nyquist distance allow to cover the visible space of spectral plane. As a result, we have the system of circles represented in Fig. 9, in which all the hexagons must belong to the visible region. As the retrieved samples have been taken at the Nyquist rate, the full field matrix can be finally obtained by using a simple Fourier interpolation.

For the first numerical example, by checking that the misfit on the unwrapped phase of trial solutions on the rings is lower than 2°, we report in Fig. 10 one of the (just) two solutions achieved at the end of the procedure[6]. As evidenced by cuts in Fig. 10 [subplot (c)], a satisfactory PR solution has been found leading to an overall $NSE_{rf} = 3.12 \cdot 10^{-5}$.

By exploiting a calculator having an Intel(R) Core(TM) i7-9700 CPU and a 32 GB RAM, the numerical reconstruction took roughly seven hours. As it can be seen, the proposed method has been able to retrieve not only the far-field phase, but also the term related to the reflector deformation [see subplots (b) and (c) of Fig. 10].

Note that the antenna at hand is four times larger than the one we considered in [2] (which was the larger dimension we were able to manage with that approach) and larger sources can be also considered in view of the different strategy.

---

[6] Hence, a further single bit of information is needed to get uniqueness, which can be eventually achieved as discussed in [2].

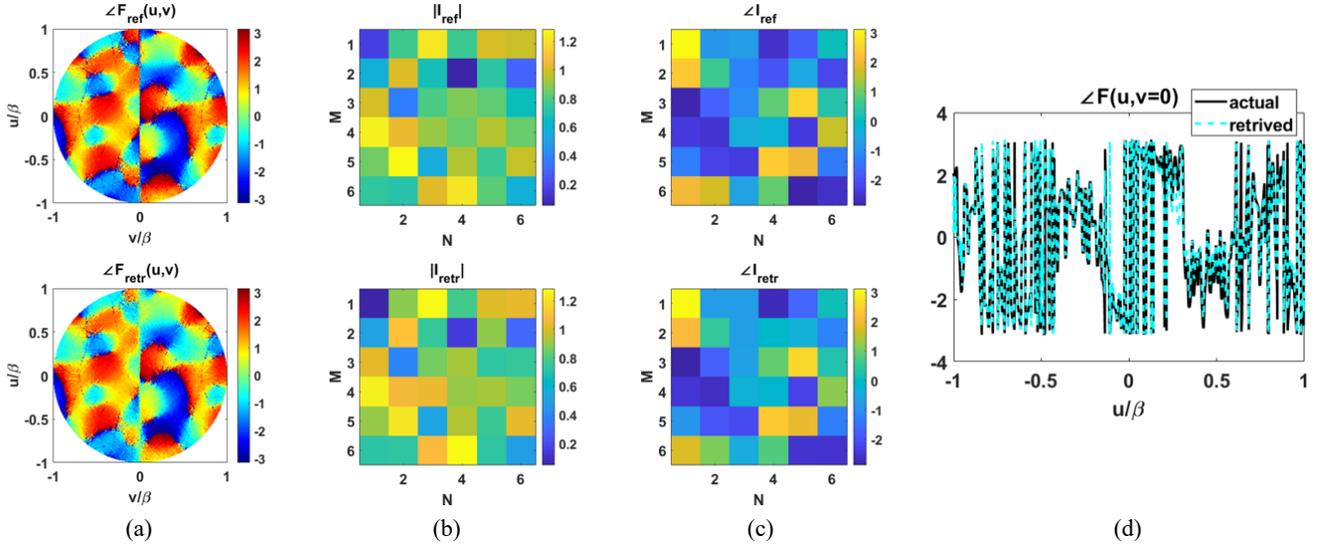

Fig. 14: Example #3: complex random excitations retrieval for a 6x6 planar array of bow-tie antennas (SNR=25dB). From left to right: (a) reference (top) and retrieved (bottom) phase of the radiated far-field; (b) reference (top) and retrieved (bottom) amplitude excitations; (c) reference (top) and retrieved (bottom) phase excitations; (d) 1-D cuts of the reference (continuous black curve) and retrieved (dashed-cyan curve) far-field phase in $v = 0$.

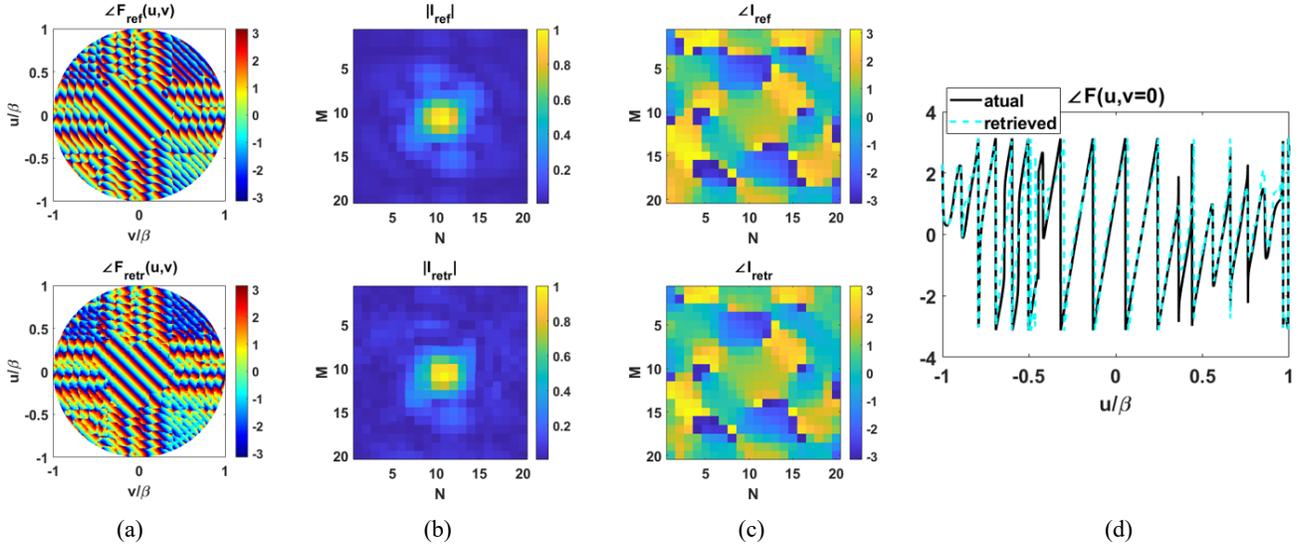

Fig. 15: Example #4: complex excitations retrieval for the 20x20 planar array dealt with in [32] (SNR=25dB). From left to right: (a) reference (top) and retrieved (bottom) phase of the radiated far-field; (b) reference (top) and retrieved (bottom) amplitude excitations; (c) reference (top) and retrieved (bottom) phase excitations; (d) 1-D cuts of the reference (continuous black curve) and retrieved (dashed-cyan curve) far-field phase in $v = 0$.

To check robustness against noise, the proposed PR strategy has been then applied to data corrupted by Gaussian white noise, where the signal-to-noise ratio (SNR) was equal to 25dB. Consequently, the 1-D PR problem solutions are affected by errors due to noise on data. Thus, the phase misfit tolerance was changed to 13°, resulting in a more difficult pruning (as more solutions become admissible). Despite this, the procedure again succeeded in finding the actual spectrum (and its complex conjugate companion) and the corresponding aperture distributions, with just a little increase (roughly 15%) in computing time.

Based on the results in Fig. 11, corresponding to $NSE_{rf} = 6.93 \cdot 10^{-4}$, it can be concluded that the proposed method is effective even when a moderate noise on data is present.

### B. Excitations Retrieval of Planar Arrays

In this Subsection we consider the retrieval of the complex excitations of 2-D array antennas by exploiting their radiation power pattern as measured data. As discussed in [4], this specific challenge represents an important PR problem in the field of microwave antenna measurements.

In all the examples below, we enforced that the misfit on the unwrapped phase of trial solutions on the rings is lower than 13°, and we corrupted each measured field amplitude with SNR=25dB.

In the first test case, to validate the applicability of the proposed approach to cases where mutual-coupling and mounting-platform effects play a role, we exploited as reference the *φ-component* of the power pattern radiated by the antenna shown in Fig. 12, i.e., a 6x6 planar array of identical bow-tie antennas [31] with *complex* random excitations in the range [-1,1] and a constant inter-element spacing equal to $0.707\lambda$ (see [30] for more details). By referring to Fig. 13, it is: L=10 mm, W=15 mm, d=2.07 mm, while the central frequency is 3 GHz.

Before executing the PR procedure, the Active Element Patterns (AEPs)[7] of the elements of the array have been computed through the CST Microwave Studio full-wave software. Then, to perform the retrieval, we used 30 concatenating rings of radius equal to half of the Nyquist distance.

One of the two solutions in terms of spectra and excitations are shown in Fig. 14 [subplot (a)] and 14 [subplots (b) and (c)], respectively. As it can be seen in Fig. 14 [subplot (d)], notwithstanding the non-regular behavior of the excitations, a satisfactory solution has been found, leading to $NSE_{rf} = 5.26 \cdot 10^{-4}$ in roughly half an hour.

As a last numerical example, we checked the procedure on a 'structured' pattern. In particular, we considered the same array and excitations as the ones in [32], i.e., a 20x20 isotropic antennas with a constant $0.5\lambda$ spacing guaranteeing a 'flat-top' footprint covering China. In such a case, we considered 648 concatenating rings of radius still equal to half of the Nyquist distance.

Reconstruction results pertaining to one of the two final outcomes are shown in Fig. 15, from which it is possible to observe and confirm the effectiveness of the proposed approach, which is also witnessed by a $NSE_{rf} = 7.46 \cdot 10^{-4}$.

## VI. CONCLUSIONS

An innovative strategy for an effective 2-D phase retrieval of radiated complex fields starting from amplitude-only measurements on a single surface has been presented and assessed.

The proposed procedure takes advantage from the 'crosswords' paradigm introduced in [1],[2], but relies herein on the intersection of curves (i.e., rings) having a much smaller length. Consequently, on these curves one has to deal with fields having a small rate of variability, which correspond to low orders of the associated polynomials, and to a small number of possible solutions along each ring. Hence, both each single factorization problem and (which is more important) the pruning of the tree of possible combinations are greatly simplified. In summary, the new choice and the associated new procedures allow us to definitively overcome the drawbacks related to computational burden of our previous approaches [1],[2] (which already resulted more effective than iterative algorithms such as the ones listed in [7]).

The overall procedure has been successfully assessed in case of sources different and considerably larger than the ones in [1],[2], including reflector and array antennas with noisy data. Notably, as opposite to almost all existing methods, it only requires a single measurement set (plus some minimal additional a-priori information able to solve a 1-bit ambiguity [2]) and hence offers definite advantages in terms of measurement time over the more standard 'two-sets-of-data' techniques. Finally, by using the 'reduced radiated field' concept [22],[23], the presented approach can also be used in the case of near-field data.

Let us finally note that, in order to trigger the procedure, one may choose a triplet of circles other than the one proposed in subsection IV.A above. Such a degree of freedom suggests a further possible optimization of the procedure. In fact, one can consider two or more clusters of rings and solve separately (i.e., in parallel, with definite computational advantages) the problem on the different clusters. In the end, a proper choice of a phase constant associated to each cluster will provide the correct concatenation amongst the different parts of the spectrum. Notably, this is indeed another similarity with crosswords puzzles solution schemes.

---

[7] The *n-th* AEP is the field radiated by the array when the *n-th* element has unitary excitation while all the other elements are closed on a matched load. It takes into account mutual-coupling and mounting-platform effects.